\documentclass[runningheads]{llncs}
\usepackage{graphicx}
\usepackage{caption}
\usepackage{subcaption}
\usepackage{amsmath}
\usepackage{amssymb}
\usepackage{bm}
\usepackage{url}
\newcommand{\etal}{\textit{et al.\,}}
\newcommand{\eg}{\textit{e.g.}}
\newcommand{\ie}{\textit{i.e.}}

\usepackage{cleveref}
\crefname{appendix}{App.\negthinspace\,}{App.\negthinspace\,}
\crefname{chapter}{Chap.\negthinspace\,}{Chap.\negthinspace\,}
\crefname{equation}{Eq.\negthinspace\,}{Eq.\negthinspace\,}
\crefname{algorithm}{Alg.\negthinspace\,}{Alg.\negthinspace\,}
\crefname{section}{Sec.\negthinspace\,}{Sec.\negthinspace\,}
\crefname{subsection}{Sec.\negthinspace\,}{Sec.\negthinspace\,}
\crefname{subsubsection}{Sec.\negthinspace\,}{Sec.\negthinspace\,}
\crefname{figure}{Fig.\negthinspace\,}{Fig.\negthinspace\,}
\crefname{table}{Tab.\negthinspace\,}{Tab.\negthinspace\,}
\crefname{subfigure}{Fig.\negthinspace\,}{Fig.\negthinspace\,}
\crefname{subsubfigure}{Fig.\negthinspace\,}{Fig.\negthinspace\,}
\crefname{lstlisting}{Lst.\negthinspace\,}{Lst.\negthinspace\,}

\begin{document}
\title{Semi- and Self-Supervised Multi-View Fusion of 3D Microscopy Images using Generative Adversarial Networks\thanks{This work was funded by the German Research Foundation DFG with the grant STE2802/2-1 (DE).}}
\titlerunning{Semi- and Self-Supervised Multi-View Fusion of 3D Microscopy Images}
%
\author{Canyu Yang\inst{1} \and Dennis Eschweiler\inst{1}\orcidID{0000-0003-1041-2810} \and Johannes Stegmaier\inst{1}\orcidID{0000-0003-4072-3759}}
%
\authorrunning{C. Yang et al.}
%
\institute{Institute of Imaging and Computer Vision, RWTH Aachen University, Aachen, Germany,
\email{johannes.stegmaier@lfb.rwth-aachen.de}}
%
\maketitle              
\begin{abstract}
Recent developments in fluorescence microscopy allow capturing high-resolution 3D images over time for living model organisms. To be able to image even large specimens, techniques like multi-view light-sheet imaging record different orientations at each time point that can then be fused into a single high-quality volume. Based on measured point spread functions (PSF), deconvolution and content fusion are able to largely revert the inevitable degradation occurring during the imaging process. Classical multi-view deconvolution and fusion methods mainly use iterative procedures and content-based averaging. Lately, Convolutional Neural Networks (CNNs) have been deployed to approach 3D single-view deconvolution microscopy, but the multi-view case waits to be studied. We investigated the efficacy of CNN-based multi-view deconvolution and fusion with two synthetic data sets that mimic developing embryos and involve either two or four complementary 3D views. Compared with classical state-of-the-art methods, the proposed semi- and self-supervised models achieve competitive and superior deconvolution and fusion quality in the two-view and quad-view cases, respectively.

\keywords{Multi-View Fusion \and Convolutional Neural Networks \and Image Deconvolution \and Multi-View Light-Sheet Microscopy.}
\end{abstract}
\section{Introduction} \label{sec:introduction}
The life science community has put forward increasing efforts in 3D+t microscopy imaging to enable detailed recordings of biological activity in living model organisms such as gene expression patterns, tissue formation and cell differentiation \cite{huisken09,multiviewimaging,Chhetri15}. Fluorescence microscopy as the most commonly used technique for observing live embryos has been facing limitations summarized by the design-space tetrahedron: resolution, speed, phototoxicity and imaging depth \cite{care}. Light-sheet microscopy achieves real-time imaging speed and lessens phototoxicity using thin optical sectioning \cite{huisken09,Chhetri15}. By recording the same specimen from multiple orientations, \eg, with multi-view light-sheet microscopy~\cite{multiviewimaging,Chhetri15}, even larger specimens can be imaged at high spatial resolution. Subsequently, multi-view fusion techniques can be used to merge the best-quality image content from all input views into a consistently sharp fusion volume~\cite{cbif}. Compared with content-based multi-view fusion~\cite{cbif}, multi-view deconvolution essentially performs fusion of deconvolved views and thus results in sharper images~\cite{ebmd}.

Inspired by the CycleGAN-based 3D single-view deconvolution microscopy by Lim \etal~\cite{lim19,lim20}, we extend their approach to 3D multi-view deconvolution and fusion using semi- and self-supervised network architectures and PSFs measured from experiments~\cite{ebmd}. Compared with the state-of-the-art method EBMD~\cite{ebmd}, the proposed models achieve comparable and superior deconvolution and fusion quality on a set of synthetic two-view and quad-view data sets, respectively.

\section{Related Work} \label{sec:relatedwork}
Deep learning has been extensively exploited to resolve single-view deblurring and deconvolution for 2D natural images~\cite{schuler13,xu2014deep,schuler15,zhang2017learning,son2017fast,wang2017deepdeblur,wang2018training,deblurgan}. Regarding 3D fluorescence microscopy images, Weigert \etal investigated and proved the efficacy of deep learning in single-view deconvolution using 2D slices~\cite{care}. Lim \etal introduced the CycleGAN architecture to 3D blind deconvolution microscopy using unpaired ground-truth images that were obtained with classical methods to regularize the image style of fusion volumes~\cite{cyclegan,lim19}. This semi-supervised model was further adapted to non-blind deconvolution by Lim \etal such that known PSFs can be explicitly implemented~\cite{lim20}. To our best knowledge, deep learning has not been introduced to multi-view deconvolution and fusion for 3D microscopy images yet, while traditional methods have delivered reasonable performance. For instance, Verneer \etal designed an optimization-based deconvolution algorithm for light-sheet microscopy images that minimizes the difference between degraded fusion volume and input views using maximum \textit{a posteriori} estimation with Gaussian noise (MAPG)~\cite{mapg}. Preibisch \etal proposed content-based image fusion (CBIF) where the fusion volume is computed as the entropy-weighted average of all input views~\cite{cbif}. In 2014, Preibisch \etal adapted the Richardson-Lucy deconvolution algorithm~\cite{richardson,lucy} to multi-view geometry and derived more efficient update schemes~\cite{ebmd}. This algorithm named Efficient Bayesian-based Multi-view Deconvolution (EBMD) achieved the state-of-the-art fusion quality compared with CBIF~\cite{cbif} and MAPG~\cite{mapg} and will serve as the baseline for benchmarking our learning-based pipelines for multi-view deconvolution and fusion that are presented in the next section.

\section{Methods} \label{sec:methods}
\begin{figure}[ht]
    \centering
    \includegraphics[width=0.9\textwidth]{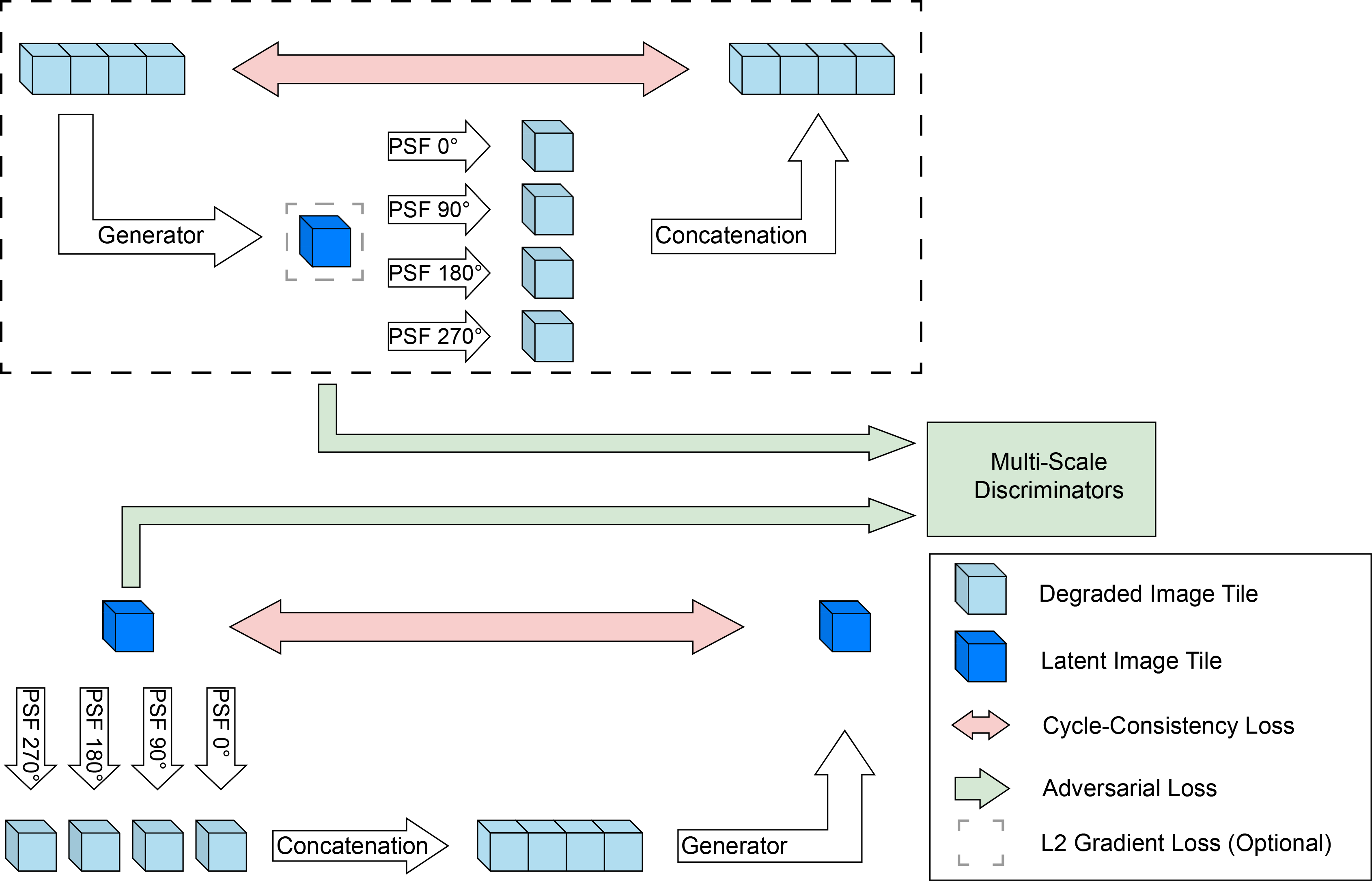}
    \caption[Semi- and self-supervised quad-view deconvolution and fusion pipeline]{Semi- and self-supervised quad-view deconvolution and fusion pipeline. Top: generator fuses four raw input views into a sharp latent image (multi-view imaging around y-axis). The generated latent image is then blurred by four known PSFs to reproduce the four degraded raw input views. Bottom: a deconvolved  ground truth image is transformed to four raw image views using the known PSFs and the generator is again used to reconstruct the deconvolved ground truth image. The loss function involves an adversarial loss with multi-scale discriminators to encourage the generator to produce realistic deconvolution results and an additional cycle consistency loss in both paths to restrict the target distribution so that the generated latent image can reproduce the input views through degradation. The self-supervised network does not involve known ground truth and features only the cycle-consistency loss and an L2 gradient loss, \emph{i.e.}, the variant involves only the part indicated by the dashed rectangle.}
    \label{fig:semipipeline}
\end{figure}

The image formation process in fluorescence microscopy can be considered as a latent image with an unknown distribution of fluorescence emitted from the specimen that is imaged through the microscope, producing a blurry image~\cite{goncharova}. Following the notation in~\cite{lim20,lim19}, we refer to the degraded image domain as $\mathcal{X}$ and the latent image domain as $\mathcal{Z}$. The imaging system can be modeled as the convolution of the latent image $\bm{z}\in\mathcal{Z}$ with the respective PSF $\bm{h}$, \emph{\ie}, $\bm{s}=\bm{z}*\bm{h}$, where $*$ denotes convolution operation. An observed image $\bm{x}\in\mathcal{X}$ results from the signal $\bm{s}$ being corrupted by noise $\bm{n}$ and can be modeled as:
\begin{equation} \label{eqn:imgformation}
\bm{x}=\bm{z}*\bm{h}+\bm{n}.
\end{equation}
Image deconvolution is an ill-posed inverse problem as an observed image $\bm{x}$ can result from numerous plausible explanations in the latent image domain $\mathcal{Z}$. In the multi-view case, the inverse problem is formulated for each view, making the restoration of the true latent image even more difficult. Our proposed semi- and self-supervised pipelines are shown in \cref{fig:semipipeline}.

 The generator $G_{\bm{\Theta}}$ maps a degraded image $\bm{x}$ to a noise-free sharp image $\bm{z}$. The discriminator $D_{\bm{\Phi}}$ distinguishes the generated sharp image $G_{\bm{\Theta}}(\bm{x})$ from true latent images $\bm{z}$. Considering non-blind deconvolution, the generator modeling the degradation process $\mathcal{Z}\rightarrow\mathcal{X}$ can be defined as a convolution with PSF $\bm{h}$. Consequently, the corresponding discriminator can be omitted to improve training efficiency. Least Squares GAN (LS-GAN) loss~\cite{lsgan} is adopted to minimize the adversarial loss with respect to the model parameters $\bm{\Phi}$ and $\bm{\Theta}$ as follows: 
\begin{equation} \label{eqn:lsganlossD}
    \min_{\bm{\Phi}}{\frac{1}{2}\mathbb{E}_{\bm{z}\sim p_{\bm{z}}(\bm{z})}[(D_{\bm{\Phi}}(\bm{z})-1)^2]+\frac{1}{2}\mathbb{E}_{\bm{x}\sim p_{\bm{x}}(\bm{x})}[D_{\bm{\Phi}}(G_{\bm{\Theta}}(\bm{x}))^2]}
\end{equation}
\begin{equation} \label{eqn:lsganlossG}
    \min_{\bm{\Theta}}{\mathbb{E}_{\bm{x}\sim p_{\bm{x}}(\bm{x})}[(D_{\bm{\Phi}}(G_{\bm{\Theta}}(\bm{x}))-1)^2]},
\end{equation}
where $p_{\bm{x}}(\bm{x})$ is the distribution of degraded image $\bm{x}$, and $p_{\bm{z}}(\bm{z})$ is distribution of true latent image. In line with \cite{lim20}, we found that the LS-GAN loss provided the most stable training convergence. 
The cycle consistency proposed in \cite{cyclegan} is imposed in both cycles, \ie, $\mathcal{X}\rightarrow\mathcal{Z}\rightarrow\mathcal{X}$ and $\mathcal{Z}\rightarrow\mathcal{X}\rightarrow\mathcal{Z}$, for the purpose of enforcing the generator to predict latent images in conformity with the view image formation defined in \cref{eqn:imgformation}. The cycle consistency loss is minimized in an $L1$ sense such that:
\begin{equation}
    L_{cycle}(\bm{\Theta})=\mathbb{E}_{\bm{x}\sim p_{\bm{x}}(\bm{x})}[\|G_{\bm{\Theta}}(\bm{x})*\bm{h}-\bm{x}\|_1]+\mathbb{E}_{\bm{z}\sim p_{\bm{z}}(\bm{z})}[\|G_{\bm{\Theta}}(\bm{z}*\bm{h})-\bm{z}\|_1].
\end{equation}
The overall objective function for the generator is summarized as:
\begin{equation} \label{eqn:overallobjective}
    \min_{\bm{\Theta}}{\mathbb{E}_{\bm{x}\sim p_{\bm{x}}(\bm{x})}[(D_{\bm{\Phi}}(G_{\bm{\Theta}}(\bm{x}))-1)^2]+\lambda L_{cycle}(\bm{\Theta})},
\end{equation}
where $\lambda$ is a weighting factor for the cycle consistency loss. The cycle consistency may not be able to sufficiently regularize the generator as there exist a number of permutations in the target distribution that maintain the cycle consistency but nevertheless exhibit inferior edge sharpness and intensity contrast. Inspired by Lim \etal~\cite{lim20}, we resort to unpaired ground-truth images to inform the generator of intensity variations and structural details desired to be reconstructed. We adapt the generator architecture from~\cite{lim20}, \ie, a three-level 3D U-Net~\cite{3dunet} with $4$ input channels, 64 feature maps in the first convolutional layer and doubling the number of feature maps after each of the two max pooling operation to a maximum of 256. Convolutional layers use $3\times3\times3$ kernels and are followed by instance normalization and LeakyReLU activation. For the upsampling path we use transposed convolutions. As the adversarial loss evaluated on original-size tiles mainly affects the restoration of low-frequency components, one can resort to multi-PatchGAN~\cite{imagetoimage} to reconstruct finer details using patches cropped from the original-size tiles. Accordingly, \cref{eqn:lsganlossD} and \cref{eqn:lsganlossG} are reformulated as:
\begin{equation} \label{eqn:semilsganlossD}
    \min_{\bm{\Phi}}{\frac{1}{2}\mathbb{E}_{\bm{z}\sim p_{\bm{z}}(\bm{z})}[\sum_{j=1}^{m}{(D_{\bm{\Phi}_j}(f_{j}(\bm{z}))-1)^2}]+\frac{1}{2}\mathbb{E}_{\bm{x}\sim p_{\bm{x}}(\bm{x})}[\sum_{j=1}^{m}{D_{\bm{\Phi}_j}(f_{j}(G_{\bm{\Theta}}(\bm{x})))^2}]},
\end{equation}
\begin{equation} \label{eqn:semilsganlossG}
    \min_{\bm{\Theta}}{\mathbb{E}_{\bm{x}\sim p_{\bm{x}}(\bm{x})}[\sum_{j=1}^{m}{(D_{\bm{\Phi}_j}(f_{j}(G_{\bm{\Theta}}(\bm{x})))-1)^2}]},
\end{equation}
where $m$ is the number of discriminators with parameters $\bm{\Phi}_j$, and the function $f_{j}(\cdot)$ crops patches of the $j$-th scale from the input image. We tested the configurations $m \in \lbrace 1,2,3 \rbrace$ and found that using two discriminators yielded the best results. All remaining results were obtained with two discriminators applied on original-size tiles ($64^3$ voxels) and half-size tiles ($32^3$ voxels).
To get rid of the dependence on ground truth images, we also tested a self-supervised variant of the semi-supervised pipeline by removing all loss parts involving unpaired ground truth, \ie, removing the adversarial loss and reducing the network to the dashed region in \cref{fig:semipipeline}. To suppress high-frequency artifacts in the generated images of the self-supervised model, an $L2$ gradient loss is employed as additional regularization in \cref{eqn:overallobjective} defined as:
\begin{equation}
    L_{gradient}=\frac{1}{n}\|\nabla\hat{\bm{z}}\|_{2}^{2},
\end{equation}
where $n$ is the number of voxels, and $\nabla\hat{\bm{z}}$ denotes the gradient of the generated latent image $\hat{\bm{z}}$.

\section{Experiments and Results} \label{sec:experimentsandresults}
\subsection{Datasets} \label{subsec:datasets}
Using the data simulation software and measured PSFs provided in \cite{ebmd}, we synthesized an isotropic quad-view data set that is reminiscent of early embryonic development. The quad-view setting was chosen as four views are not too memory expensive and provide the network with sufficient information. The data set consists of 140 sample groups, each of which comprises a quadruplet of view images and the associated ground truth. The training, validation and test sets contain 108, 21 and 11 sample groups, respectively. The view and ground truth images are of size $256\times256\times256$. In the following context, this data set is referred to as embryo data set. Based on 3D images of \textit{C. elegans} after worm body straightening from~\cite{nuclei}, we simulated a two-view data set resembling the nuclei distribution of \textit{C. elegans} using PSFs provided in~\cite{ebmd}. Compared with the embryo data set, the simulated nuclei are more sparsely-distributed and have fewer touching boundaries. This nuclei data set contains 80 sample groups, among which 68 groups are taken as training set, and 12 groups as validation and test set. The raw and ground truth images are of size $140\times140\times1000$ voxels.

\subsection{Existing Methods for Comparison} \label{subsec:comparisonmethods}
To evaluate the performance of our proposed multi-view deconvolution and fusion pipelines, we compare it to an uncorrected raw image (View 0°), the CBIF \cite{cbif} and the EBMD~\cite{ebmd} algorithms. Both methods are the original Fiji plug-in implementations by the respective authors~\cite{fiji,spimregistration,ebmd}. The basic form of EBMD was utilized as this variant provided higher accuracy than the variants with type-I or type-II optimization. For the embryo data set, we adopted $48$ for the number of iterations and $0.004$ for the weighting factor of the Tikhonov regularization~\cite{tikhonov} as recommended in~\cite{ebmd}. In the context of the nuclei data set, the optimal hyperparameters were manually tuned and empirically set to 15 iterations and a weighting factor of $0.1$.

Normalized root-mean-square error (NRMSE)~\cite{care}, peak signal-noise-ratio (PSNR), structural similarity index (SSIM)~\cite{ssim} and the correlation coefficient (CC) were adopted to quantify the image quality of the multi-view deconvolution results. All metrics were evaluated with MATLAB on the test set of each data set. As the deconvolution results from different methods vary in their value range, it is required to normalize the results prior to quantification. We adopted the normalization method by Weigert \etal~\cite{care}, to cope with varying absolute intensity levels of the different methods. To be consistent throughout metric evaluation, we utilized the same percentile-normalized ground-truth images for the evaluation of other metrics. Importantly, we evaluated each metric in two different settings: one with all voxels, and the other one with only the foreground voxels, \ie, only the voxel positions with non-zero intensity values in the normalized ground truth are considered. The rationale is to measure the deconvolution capability of the proposed models in terms of the content of interest since artifacts in the empty background can interfere with the metric evaluation.

\begin{figure}[htb]
    \centering
    \includegraphics[width=\textwidth]{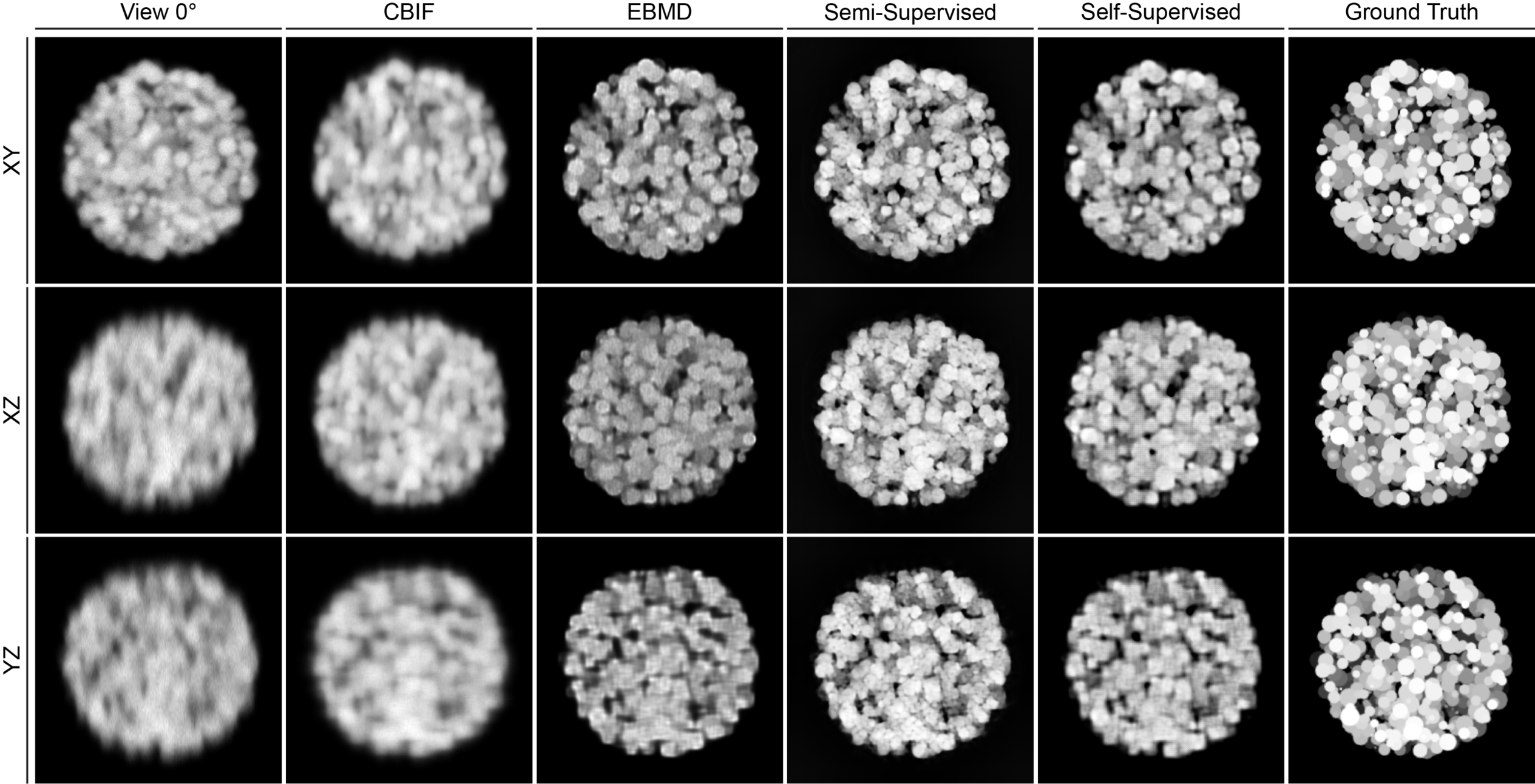}
    \caption[Deconvolution and fusion quality comparison between the proposed models and the existing methods using the embryo data set]{Deconvolution and fusion quality comparison on the embryo data set. The columns show an uncorrected raw image, the results of the CBIF, EBMD, Semi-Supervised, Self-Supervised methods and the ground truth (left to right).}
    \label{fig:quadviewresults}
\end{figure}
\subsection{CNN-based Multi-View Deconvolution and Fusion} \label{subsec:semiandselfresults}
All models were implemented in Pytorch and code is available from \url{https://github.com/stegmaierj/MultiViewFusion}. Given the embryo data set, both the semi- and self-supervised models were trained with input volumes of size $4\times64\times64\times64$, batch size of 1, learning rate of 0.0001, 90 epochs, cycle loss weight set to $\lambda=10$ using the Adam optimizer~\cite{adam}. Due to the requirement of the semi-supervised pipeline for unmatched ground truth, the training set in the semi-supervised case was equally divided into two subsets: one half was utilized as inputs to the network while the other half served as provider for unmatched ground truth. The deconvolution and fusion results of the two proposed models are displayed and compared with the existing methods in \cref{fig:quadviewresults}.
Compared with EBMD~\cite{ebmd}, our methods produce commensurate sharpness and structural details. Furthermore, both proposed pipelines yield stronger brightness contrast in the $XY$- and $XZ$-slices and remove more blur in the $YZ$-slices.
Based on the performance metrics listed in \cref{tab:quadview}, our methods outperform EBMD~\cite{ebmd} regarding both all-voxel and foreground-only evaluations. The semi-supervised model gives an inferior SSIM evaluated on all voxels, which presumably arises from the amplification of noisy voxels in the background due to the intensity normalization in~\cite{care}. As SSIM is evaluated in a sliding-window manner it is susceptible to local variations caused by the falsely amplified background intensities. This is corroborated by the SSIM evaluated only with foreground voxels.
\begin{table}[tb]
    \centering
    \caption{Deconvolution and fusion quality evaluated on the embryo (top) and the nuclei data set (bottom). Left subcolumn: background and foreground, right subcolumn: only foreground.}
    \label{tab:quadview}
    \begin{tabular}{|l|l|l|l|l|l|l|l|l|}
        \hline
        Methods & \multicolumn{2}{|l|}{NRMSE} & \multicolumn{2}{|l|}{PSNR(dB)} & \multicolumn{2}{|l|}{SSIM} & \multicolumn{2}{|l|}{CC}\\
        \hline
        View $0^{\circ}$ & 0.104 & 0.185 & 19.685 & 14.645 & 0.546 & 0.845 & 0.949 & 0.735\\
        CBIF~\cite{cbif} & 0.101 & 0.181 & 19.913 & 14.857 & 0.508 & 0.852 & 0.951 & 0.749\\
        EBMD~\cite{ebmd} & 0.078 & 0.145 & 22.182 & 16.763 & 0.838 & 0.885 & 0.971 & 0.847\\
        Semi-supervised & 0.063 & 0.111 & 23.998 & 19.089 & 0.615 & 0.913 & 0.981 & 0.914\\
        Self-supervised & \textbf{0.057} & \textbf{0.107} & \textbf{24.825} & \textbf{19.429} & \textbf{0.900} & \textbf{0.915} & \textbf{0.985} & \textbf{0.920}\\
        \hline
        \hline
        View $0^{\circ}$ & 0.084 & 0.278 & 21.549 & 11.151 & 0.488 & 0.948 & 0.862 & 0.638\\
        CBIF~\cite{cbif} & 0.079 & 0.273 & 22.050 & 11.328 & 0.614 & 0.950 & 0.878 & 0.656\\
        EBMD~\cite{ebmd} & \textbf{0.079} & 0.268 & \textbf{22.118} & 11.460 & 0.634 & 0.952 & \textbf{0.881} & 0.670\\
        Semi-supervised & 0.087 & \textbf{0.262} & 21.245 & \textbf{11.655} & 0.602 & \textbf{0.955} & 0.852 & \textbf{0.688}\\
        Self-supervised & 0.082 & 0.276 & 21.767 & 11.212 & \textbf{0.674} & 0.954 & 0.869 & 0.645\\
        \hline
    \end{tabular}
\end{table}
As for the two-view nuclei data set, the training settings remain the same except for an input volume size of $2\times16\times128\times960$ and training for 500 epochs (other tile sizes like $64^3$ and $128^3$ consistently performed worse). Similarly, the training set for semi-supervised learning was equally divided into two subsets.

\begin{figure}[htb]
    \includegraphics[width=\textwidth]{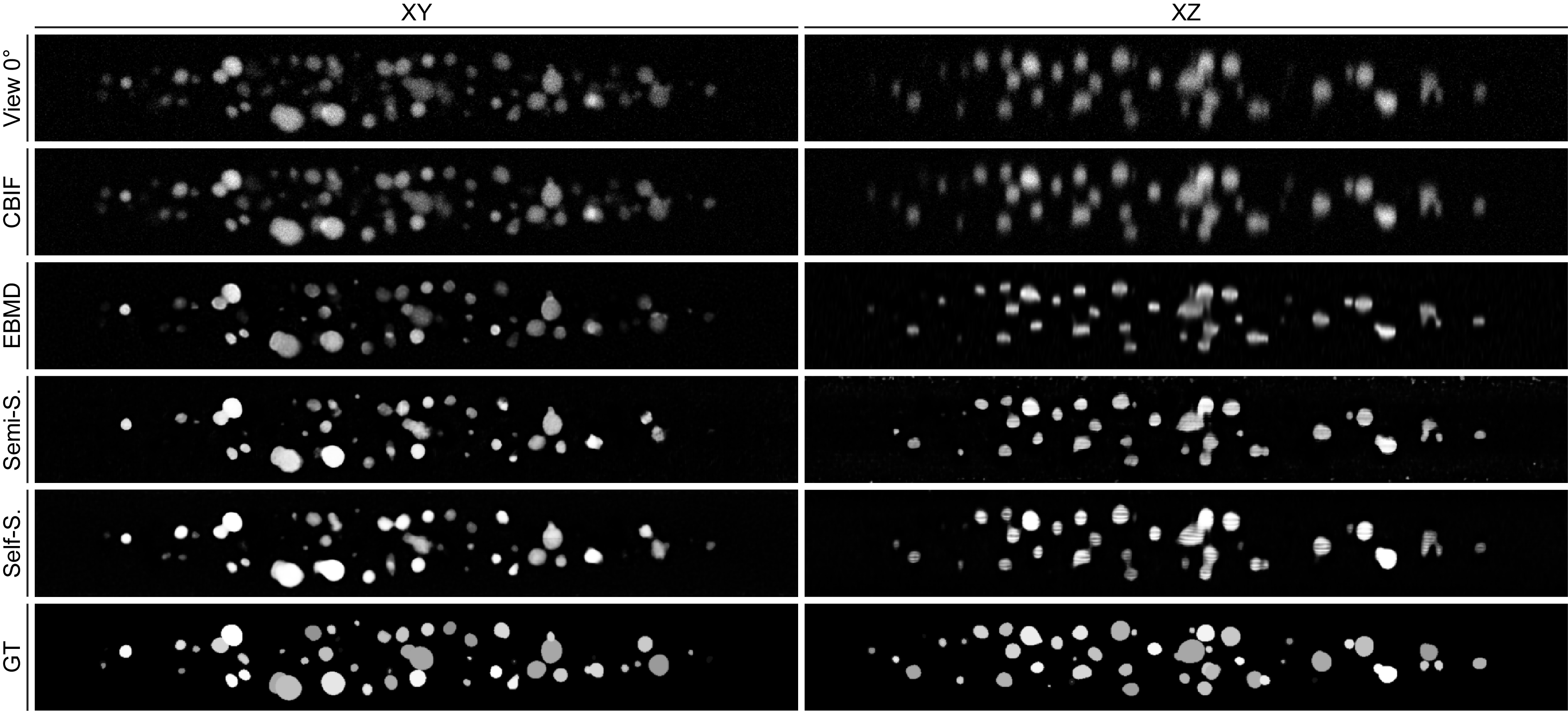}
    \caption{Deconvolution and fusion quality comparison on the nuclei data set for corresponding XY slices (left column) and XZ slices (right column). The rows show an uncorrected raw image, the results of the CBIF, EBMD, Semi-Supervised, Self-Supervised methods and the ground truth (top to bottom).}
    \label{fig:twoviewresults}    
\end{figure}

Qualitatively our methods produce deblurring quality slightly superior to EBMD~\cite{ebmd} (\cref{fig:twoviewresults}). However, in the $XZ$-slices high-frequency artifacts are generated by the GAN-based models. As shown in \cref{fig:twoviewresults}, the self-supervised model suffers from the high-frequency artifacts in nuclei areas. EBMD~\cite{ebmd} outperforms the semi-supervised model for all-voxel evaluation as the proposed model generates artifacts in the dark background, while given only foreground content our method yields the best quality.


\section{Conclusions} \label{subsec:discussionandconclusion}
In this work we approached 3D non-blind multi-view deconvolution and fusion using two CNN-based pipelines on synthetic microscopy data sets. In the quad-view case, both the semi- and self-supervised pipelines outperform the state-of-the-art method EBMD~\cite{ebmd} quantitatively and qualitatively in terms of sharpness and crispiness.With respect to the ground truth, the self-supervised model delivered the most reasonable deblurring quality and brightness contrast. Given the two-view data sets, the semi-supervised model surpasses EBMD~\cite{ebmd} regarding foreground content while the self-supervised model suffers from the high-frequency artifacts in objects of interest. Consequently, the semi-supervised pipeline is more adaptive in terms of the number of input views and the distribution density of objects of interest. Without using any ground truth, the self-supervised pipeline has demonstrated to be a viable alternative to classical methods. A limitation of the method is that the PSFs are assumed to be known and are kept constant during the training of the GAN. For a more generally applicable method, conditioning the GAN with arbitrary PSFs or performing a blind deconvolution would be interesting next steps. To suppress the high-frequency artifacts, more prior knowledge of the true latent image would be required to regularize the latent image prediction. We intentionally focused on synthetic data in this study that allowed us to quantitatively assess the deconvolution and fusion quality of the different approaches. A logical next step will be to assess how the proposed methods translate to real microscopy images. Moreover, in future work we will also investigate how deeper network architectures, different patch sizes and voxel-shuffle or other upsampling methods can potentially be used for artifact removal in the results of the self-supervised model. 
%
%
%

\bibliographystyle{splncs04}
\bibliography{references}
%




\end{document}